\documentclass{amsart}

\usepackage{amsmath, graphicx, amssymb, mathrsfs}
\input xy
\xyoption{all}
\usepackage{amsfonts}
\usepackage{amsthm}

\swapnumbers
\newtheorem{theorem}{Theorem\setcounter{nitem}{0}}[section]
\newtheorem{definition}[theorem]{Definition\setcounter{nitem}{0}}

\newtheorem{lemma}[theorem]{Lemma\setcounter{nitem}{0}}
\newtheorem{nitem}{}

\newtheorem{eg}[theorem]{Example\setcounter{nitem}{0}}
\newcommand{\R}{\mathbb{R}}
\newcommand{\TT}{\mathbb{T}}
\newcommand{\NN}{\mathbb{N}}
\newcommand{\cal}{\mathcal}
\newcommand{\comment}[1]{}

\newcommand{\PP}{\mathbb{P}}
\newcommand{\MM}{\mathbb{M}}

\newcommand{\J}{{\cal J}}

\newcommand{\N}{\mathbb{N}}

\newcommand{\C}{\mathbb{C}}

\newcommand{\dd}{\partial}
\newcommand{\db}{\bar{\partial}_b}
\newcommand{\im}{\Im}

\title[Null electromagnetic fields]{Null electromagnetic fields and relative CR embeddings}

\author[J.E. Holland]{Jonathan Earl Holland}
\author[G. Sparling]{George Sparling}
\address{Mathematics Department\\ 301 Thackeray Hall\\ University of Pittsburgh\\ Pittsburgh, PA \ \ 15260}
\email{jeh89@pitt.edu}

\begin{document}
%\nocite{*}
\maketitle
\begin{abstract}
This paper applies the notion of relative CR embeddings to study two related questions.  First, it answers negatively the question posed by Penrose whether every shear-free null rotating congruence is analytic.  Second, it proves that given any shear-free null rotating congruence in Minkowski space, there exists a null electromagnetic field which is null with respect to the given congruence.  In the course of answering these questions, we introduce some new techniques for studying null electromagnetic fields and shear-free congruences in general based on the notion of a relative CR embedding.
\end{abstract}
%\keywords{CR manifolds; CR embeddings; null electromagnetic fields; shear-free congruences; twistor theory}
%Please type here List of Keywords for your article separated by semicolon.

%\subjclass{32V30; 32V15; 83C50; 83C60} % e.g. 35A30; 81Q05

\section{Introduction}\label{intro}

A null geodesic congruence $\cal F$ in (signature $(1,3)$) Minkowski space $\MM$ is a foliation of an open set in Minkowski space by oriented null geodesics.  Such a foliation generates a tangent vector field $k^a$.  The congruence $\cal F$ is said to be {\it shear-free} if there exists a complex null vector field $l^a$ (called a {\it connecting vector} for the congruence) such that 
$$[k,l]^a=0$$
$$k^al_a=0$$
$$k^{[a} l^b\bar{l}^{c]}\not=0.$$ 
The congruence is {\it rotating} if in addition
$$l^{[a}\bar{l}^b[l,\bar{l}]^{c]}\not= 0.$$
(Shear and rotation will be defined in Section \ref{congruences} as certain Frobenius obstructions.)  Shearfree congruences were introduced in Robinson \cite{Robinson} and linked to the study of null solutions of the electromagnetic fields.  They remain an important tool in the study of algebraically special solutions of Einstein's equations (\cite{RobinsonTrautman1960}, \cite{RobinsonTrautman1962}, \cite{KerrSchild}, \cite{Sommers}).  A detailed history can be found in Trautman \cite{Trautman}.

An electromagnetic field is a 2-form $F$ such that
$$dF=0,\ \ d\star\!F=0$$
where $*$ is the duality operator.  A nonzero field $F$ is said to be {\em null} if
$$F_{ab}F^{ab}=F_{ab}(\star\!F)^{ab}=0,$$
where $F=F_{ab}dx^a\wedge dx^b$ and $F_{ab}=-F_{ba}$.  One can show that a field $F$ is null if and only if the energy-momentum tensor
$$T_{ac}=-{F_a}^bF_{cb}+{1\over 4}g_{ab}F^{cd}F_{cd}$$
splits as an outer product
$$T_{ac}=k_ak_c$$
for some nonzero null vector $k^a$, the principal null direction of the field.  If $F$ is a null electromagnetic field, then $k^a$ is the tangent vector for some shear-free congruence (Lemma \ref{TypeNShearfree}).  In that case, $F$ is said to be {\em adapted to $k^a$}.

This paper answers two distinct but related questions.  First, it establishes a foundation for the theory of shear-free congruences of geodesics in Minkowski space.  We prove that a given shear-free congruence corresponds to a certain CR submanifold $N$ of the indefinite hyperquadric of $\C\PP^3$ (projective twistor space in this context).  Kerr \cite{Kerr} (cf. also Penrose \cite{PenroseAlg}) has shown that the aforementioned result holds in the case of analytic congruences, where the shear-free congruence may be represented as a holomorphic surface in $\C\PP^3$; the CR manifold $N$ is then the intersection of this holomorphic surface with the hyperquadric.  In this paper, we drop the assumption of analyticity and work in the smooth setting and show that one still obtains this CR manifold $N$ (Theorem \ref{SF=>CR}).  Conversely, given a CR submanifold $N$ of the hyperquadric (which obeys some genericity criteria) one obtains a shear-free congruence (Theorem \ref{CR=>SF}).  In the smooth category, one no longer necessarily obtains the whole holomorphic surface associated with $N$ (Theorem \ref{antirob}); however, we prove here that there is a unique complex 2-surface $Z$ {\it with boundary} of which $N$ is the boundary.  The key notion here is that of a {\it relative embedding} of CR manifolds, defined in Section \ref{onesided} below.  

It is significant that in general $N$ need not be real analytic and yet corresponds to a shear-free {\em rotating} congruence.  We construct such a non-analytic $N$, contrary to the observation in Penrose \cite{PenroseAlg}: ``These [non-analytic] exceptional cases appear to occur only when the rotation also vanishes.''  The assumption of analyticity has since entered the lore of the subject, particularly in regards to the complex worldline approach of Newman and others \cite{Newman}.  Quite recently, nonanalytic congruences have been studied in Baird and Eastwood \cite{Eastwood}, although the congruences there have zero rotation.  Nonanalytic congruences with zero rotation are readily constructed from the null rays emanating from a nonanalytic worldline in real Minkowski space.  It is substantially more difficult to construct nonanalytic congruences with nonzero rotation, which is one aim of this paper.

Second, the paper applies these results to a class of solutions of the electromagnetic field equations, addressing the following question: given a shear-free rotating congruence $k^a$ what is the space of solutions of the field equations which are adapted to $k^a$?  We prove that the space of solutions corresponds to the space of all holomorphic sections of the canonical bundle of the complex surface (with boundary) $Z$ in a neighborhood of the boundary, or alternatively to $(2,0)$-forms on the CR manifold $N$ which are $\db$-closed (Lemma \ref{max=db}, Theorem \ref{localexistence}).  A special case of this result is the existence of local solutions to the electromagnetic field equations adapted to $k^a$.  This special case is discussed in Tafel \cite{Tafel} using rather different techniques.  

\section{CR manifolds}\label{CRManifolds}
Let $M$ be a smooth real manifold of dimension $2n+k$, where $n$ and $k$ are fixed integers.  A CR structure on $M$ of type $(n,k)$ is an involutive distribution $D$ of complex $n$-planes on $M$ such that $D\cap\bar{D}=\{0\}$.  The pair $(M,D)$ is called a {\it CR manifold}.  The number $n$ is called the CR dimension of $M$ and $k$ the CR codimension.  A CR manifold for which $k=1$ is said to be of hypersurface type.  By abuse of language, we often omit the $D$, and call $M$ itself a CR manifold.  In this case, when we wish to refer explicitly to the CR structure, we define $T^{1,0}M\overset{\text{def}}{=}D$ and $T^{0,1}M\overset{\text{def}}{=}\bar{D}$.

Given a CR manifold $(M,T^{1,0}M)$, it is often convenient to have a description of the CR structure exclusively in terms of complex 1-forms on $M$.  Let $\Omega^{0,1}M$ be the subsheaf of $\Omega^1M\otimes\C$ which annihilates $\Gamma(T^{1,0}M)$, and $\Omega^{1,0}M=\overline{\Omega^{0,1}M}$.  The ideal $\cal J$ in $\Omega M\otimes\C$ generated by $\Omega^{0,1}M$ is a differential ideal:
$$d{\cal J}\subset {\cal J}.$$
(Actually, the condition that $d{\cal J}\subset {\cal J}$ is necessary and sufficient for $T^{1,0}M$ to be involutive.)  

\begin{eg}  A complex manifold $M$ inherits the structure of a CR manifold of CR codimension $0$ by letting $T^{1,0}M$ be the holomorphic tangent bundle.  Conversely, the Newlander--Nirenberg theorem \cite{nn} asserts that every codimension $0$ CR manifold arises in this way.
\end{eg}

\begin{eg}  A real hypersurface $M$ in a complex manifold $Y$ inherits the structure of a CR manifold of CR codimension $1$ by setting
$$T^{1,0}M=(TM\otimes\C)\cap (T^{1,0}Y)|_M.$$
\end{eg}

The {\it tangential Cauchy-Riemann complex} is defined as follows in Hill and Nacinovich \cite{hn6}.  Let 
$$Q^{0,1}(M)= (\Omega^1(M)\otimes\C)/\Omega^{1,0}(M).$$  Then $Q^{0,1}(M)$ is the dual space of $T^{0,1}$.  For $f\in C^\infty(M)\otimes\C$, define $\db f\in Q^{0,1}(M)$ to be the coset of $df$ modulo $\Omega^{1,0}(M)$.  More generally, since $d\J\subset\J$, we have $d\J^k\subset\J^k$ for all integers $k\ge 0$ (where $\J^0:=\Omega(M)\otimes\C$).  Upon passing to the quotient, $d$ induces a mapping
$$\tilde{d}:\J^k/\J^{k+1}\rightarrow\J^k/\J^{k+1}.$$  Let $Q^{p,j}(M)$ be the graded part of $\J^p/\J^{p+1}$ of degree $(p+j)$.  Thus 
$$Q^{p,j}(M)=(\J^p\cap\Omega^{p+j}(M)\otimes\C)/(\J^{p+1}\cap \Omega^{p+j}(M)\otimes\C)$$
by definition.  Note that 
$$\tilde{d}|_{Q^{p,j}}:Q^{p,j}(M)\rightarrow Q^{p,j+1}(M)$$ 
since $d$ is a graded differential of degree $+1$.  We write $\tilde{d}=\db$ and call $\db$ the tangential Cauchy-Riemann operator.  The tangential Cauchy-Riemann (or just CR) complex is then the complex
$$0\rightarrow Q^{p,0}(M)\xrightarrow{\db}Q^{p,1}\xrightarrow{\db}...$$
where by convention 
$$Q^{0,0}=C^\infty(M)\otimes\C.$$  
Let $Q$ be the bigraded module $Q=\bigoplus_{p,q} Q^{p,q}$.  This is indeed a complex, for $\db\db$ is given as the composition $dd=0$ modulo a suitable ideal.  When a distinction is needed in the action of $\db$ on the $Q^{p,j}$, denote by $\db^{p,j}$ the (graded) part of the coboundary which maps $Q^{p,j}\rightarrow Q^{p,j+1}$.

Let $M$ be a CR manifold of hypersurface type.  The {\it Levi form} is the hermitian form 
\begin{equation*}
\begin{aligned}
{\cal V}:(T^{1,0}(M)\oplus T^{0,1}(M))\times(T^{1,0}(M)\oplus &T^{0,1}(M))\rightarrow\\
&\rightarrow {T(M)\otimes\C\over T^{1,0}M\oplus T^{0,1}M},
\end{aligned}
\end{equation*}
defined on sections $v,w$ by
$${\cal V}(v,w)={1\over 2i} [v,\bar{w}]\mod{ T^{1,0}M\oplus T^{0,1}M}.$$
Note that $\cal V$ behaves as a tensor under change in basis sections, and therefore represents a well-defined bundle map.  If the Levi form is nondegenerate and has definite signature, then $M$ is called strictly pseudoconvex.

If $M$ and $N$ are two CR manifolds, then a morphism of CR structures is a smooth function $f:M\rightarrow N$ such that
$$f_*T^{1,0}M\subset T^{1,0}N.$$
A morphism is an {\it embedding} if it is a smooth embedding in the sense of differential topology.

\section{One-sided embeddings of CR manifolds}\label{onesided}
The following definition is due to Hill, \cite{hn2}:

\begin{definition}  A complex manifold with abstract boundary is a real $2n$-dimensional manifold $M$ with boundary, together with an involutive distribution of complex $n$-planes on $M$, denoted $T^{1,0}M$, such that $T^{1,0}M\cap \overline{T^{1,0}M}=0$.  
\end{definition}

In the above definition, $T^{1,0}M$ will be called the holomorphic structure of $M$.  Let $\Omega^{0,1}M$ be the sheaf of annihilators of $T^{1,0}M$.  A mapping $f:M\rightarrow N$ between two complex manifolds with boundary is {\it holomorphic} if $f_*T^{1,0}M\subset T^{1,0}N$.  If $M$ is a manifold with boundary, let $\dd M$ be the boundary of $M$ and
$$M^o=M-\dd M$$
be the interior of $M$.  Note well that $M$ is a complex manifold by the Newlander--Nirenberg theorem, and that the tangential part of $T^{1,0}M$ along $\dd M$ determines a CR structure $T^{1,0}(\dd M)$, so that the boundary of a complex manifold with boundary is a CR manifold of hypersurface type..

\begin{definition} A complex manifold with concrete boundary is a complex manifold with abstract boundary which is locally isomorphic to some domain with boundary in $\C^n$.
\end{definition}

Hill \cite{hn3} proves the inequivalence of these two definitions.  For the rest of this paper, ``manifold with boundary'' shall mean ``manifold with concrete boundary.''

\begin{definition} A holomorphic vector bundle on a complex manifold $M$ with boundary is a complex vector bundle $E$ on $M$ with holomorphic structure $T^{1,0}E$ such that\\

1.  $x\oplus y\mapsto x+y$, $(\lambda, y)\rightarrow \lambda y$ are holomorphic maps from $E\oplus E\rightarrow E$ and $\C\times E\rightarrow E$, respectively.\\

2.  The projection mapping of the bundle $\pi:E\rightarrow M$ is holomorphic.\\
\end{definition}

\begin{definition} Let $N$ be a CR manifold of hypersurface type and $Y$ a complex manifold with boundary.  Let $f:N\rightarrow Y$ be an embedding of real manifolds with boundary such that $f(N)\subset\partial Y$ and $f^*(\Omega^{1,0}(\partial Y))\subset\Omega^{1,0}(N)$.  Then $f$ is said to be a {\it one-sided embedding} of $N$ in $Y$.
\end{definition}

If $X$ is a complex manifold, and $N$ is a (real) hypersurface in $X$ with 
$f:N\rightarrow X$ the inclusion, then $f$ is also a local one-sided embedding of $N$ in the following sense.  For each $p\in N$, there exists an open neighborhood $U$ of $p$ in $X$ such that $U-N$ has two components, call them $U^{+o}$ and $U^{-o}$.  Let $U^+$ and $U^-$ denote the closures of $U^{+o}$ and $U^{-o}$ in $U$.  Then $U^+$ and $U^-$ are complex manifolds with boundary $N\cap U$, and 
$f^{\pm}:N\cap U\rightarrow U^{\pm}$ are CR embeddings. 

\begin{definition} Let $Y$ be an $n+1$-dimensional complex manifold, $Q$ an embedded CR submanifold of $Y$ of real dimension $2n+1$, and $N$ a CR manifold embedded in $Q$.  

1.  The triple $(Y,Q,N)$ is called a two-sided (resp. one-sided) relative embedding if there exists a complex submanifold (resp. complex submanifold with boundary) $Z$ of $Y$ such that $N=Z\cap Q$.  We call $Z$ the two-sided (resp. one-sided) extension of $N$ in $Y$.

2.  The triple $(Y,Q,N)$ is called an essential one-sided relative embedding if it is a one-sided relative embedding which is not a two-sided relative embedding.
\end{definition}

\section{An essential one-sided embedding}\label{essentialsection}
\begin{figure}
\center\includegraphics[scale=0.7]{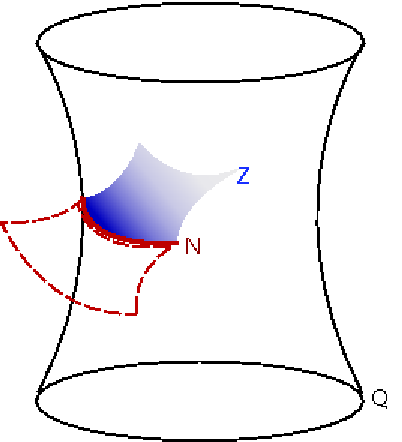}
\caption{The submanifold $N$ is two-sided embedded in $Q$ if it can be realized as the intersection of a complex surface with $Q$.  It is (essentially) one-sided embedded if the complex surface exists only on one side.}
\end{figure}
Let $\mathbb{CP}^3$ denote the complex projective $3$-space, equipped with projective coordinates $(z,v,u,w)$.  The aim of this section is to prove the following:
\begin{theorem}\label{essential}  Let $Q'$ be the hyperquadric $|z|^2+|v|^2-|u|^2-|w|^2=0$ in $\C\PP^3$.  There exists an essential one-sided relative embedding into $Q'$.
\end{theorem}

For convenience, let us work on the affine subset of $\C\PP^3$ given by $v\not= 0$, and normalize coordinates so that $v=1$.  Let $Q$ be the locus of points $(u,w,z)$ such that
$$\rho(u,w,z)\overset{\text{def}}{=}1+|z|^2-|u|^2-|w|^2=0.$$  

The CR submanifold $N$ of $Q$ shall be given as a graph
$$N=\{(u,w,z):z=wg(u)\},$$
where $g$ is a function defined on the whole complex plane.  We impose the following conditions on $g$:
\begin{nitem}  The domain of analyticity of $g$ is the complex plane slit from $1$ to $\infty$ along the positive real axis: $\C-[1,\infty)$.
\end{nitem}
\begin{nitem} $g(u)$ is continuous at $u=1$.
\end{nitem}
\begin{nitem}  $g(u)$ vanishes to infinite order at $u=1$ in the sense that
$$\lim_{u\rightarrow 1} {g(u)\over (u-1)^k}=0$$
for every positive integer $k$.  Furthermore, the restriction of $g$ to $\C-(1,\infty)$ is $C^\infty$ at $u=1$.
\end{nitem}
\begin{nitem}  $g(u)$ does not vanish away from $1$.
\end{nitem}
\begin{nitem} $|g(u)|\le {1\over 3}$ on its domain.
\end{nitem}
An example of $g(u)$ satisfying these requirements is the following:
$$g(u)=\begin{cases}
{1\over 3}\exp\left(-(1-u)^{-1/4}\right)& u\not=1\\
0&u=1
\end{cases}$$
along the branch of the logarithm for which $-\pi\le\arg (1-u) < \pi$.  Then $g(u)$ is analytic at every point of the slitted plane and continuous at $u=1$.  Furthermore, writing $1-u=re^{i\theta}$ with $r\ge 0$ and $\theta\in[-\pi,\pi)$, it follows that
$$|g(u)|= {1\over 3}\exp\left(-r^{-{1\over 4}}\cos{\theta\over 4}\right),$$
so that evidently
$$\lim_{u\rightarrow 1} {g(u)\over (u-1)^k}=0$$
and moreover, since $\cos{\theta\over 4}\ge{1\over\sqrt{2}}$ on $-\pi\le\theta<\pi$,
$$|g(u)|^2\le {1\over 9}\exp\biggl(-\left({r\over 4}\right)^{-{1\over 4}}\biggr)\le {1\over 9},$$
whence $|g(u)|\le 1/3$.

Let $E$ be the closed region in $\C^2$ defined by $2|u|^2+|w|^2\le 2$.  Let
$$t(u,w)=1+|wg(u)|^2-|u|^2-|w|^2,$$
and let $T=\{(u,w)\in E : t(u,w)=0\}$.  Let $N'$ be the graph of $z=wg(u)$ over $T$; thus
$$N'=\{(u,w,wg(u)): (u,w)\in E, t(u,w)=0\}.$$

\begin{lemma}\label{submanifold} There is a neighborhood of $U$ of $(u,w,z)=(1,0,0)$ such that $N'\cap U$ is a smooth submanifold of $Q$ containing the point $(u,w,z)=(1,0,0)$.
\end{lemma}
\begin{proof}  The fact that $N'\cap U$ contains the point $(1,0,0)$ is immediately verified from the definition.  Since $N'$ is a smooth graph over $T$, it suffices to show that $T$ is a submanifold with boundary of $E$ in a neighborhood the point $(u,w)=(1,0)$.  On the boundary of $E$,
$$t(u,w)=|w|^2\left(|g(u)|^2-{1\over 2}\right)\le 0,$$
with equality if and only if $|u|=1$ and $w=0$, by the conditions on $g$ and by the definition of the region $E$.  At the point $q=(1,0)$ we have (by the conditions on $g$ again):
$$dt(q)=-du-d\bar{u}.$$
Hence, for any vector $v$ at $q$ pointing towards the interior of $E$, we have
$$i_v dt(q)>0.$$
The restriction of $t$ to $E$ is $C^\infty$.  By the Whitney extension theorem, choose a $C^\infty$ extension $h$ of $t|_E$ to all of $\C^2$.  It follows by the implicit function theorem that the zero locus of $h$ in a sufficiently small neighborhood of $q$ is a submanifold of that neighborhood.  Now $dh=dt$ on $E$, and since $i_v dh(q)>0$ for interior-pointing vectors at $q$, and $t\le 0$ on the boundary of $E$, it follows that $t<0$ for points outside $E$ sufficiently near $q$.  Hence portion of the zero locus of $h$ in a sufficiently small neighborhood of $q$ lies entirely within $E$.  Moreover, within that neighborhood, iit touches the boundary only at $q$.
\end{proof}

Now let $N=N'\cap U$ as in Lemma \ref{submanifold}.

\begin{lemma}\label{proofofessential}  The triple $(\C^3,Q,N)$ is an essential one-sided relative embedding.
\end{lemma}

\begin{proof}
\setcounter{nitem}{0}
\begin{nitem}  $T$ is strictly pseudoconvex.
\end{nitem}

The Levi form of $T$ is represented by
\begin{equation*}
\begin{aligned}
\dd\bar{\dd}t&=\\
&\bigl(|g(u)|^2-1\bigr)dw\wedge d\bar{w} +\\
&+2i\im(g^\prime(u)g(\bar{u}-1)wdu\wedge d\bar{w})+\\
&+\bigl(|g^\prime(u)|^2|w|^2-1\bigr)du\wedge d\bar{u}.
\end{aligned}
\end{equation*}
As $u\rightarrow 1$, $g(u),g'(u)\rightarrow 0$, so this is a negative definite form in a sufficiently small neighborhood of $(u,w)=(1,0)$, and so $T$ is strictly pseudoconvex.

\begin{nitem}  The smooth embedding of $T$ in $Q$ given by
$$j:(u,w)\mapsto (u,w,wg(u))$$
is a CR embedding.
\end{nitem}

The CR structure $T^{1,0}(T)$ is generated by the vector field
$$L=\bigl(|w|^2g(\bar{u}-1)g^\prime(u)-\bar{u}\bigr){\dd\over\dd w}-\bar{w}\bigl(|g(u)|^2-1\bigr){\dd\over\dd u}.$$
Note in particular that $L$ annihilates $t$.  Now, since the function $j$ is holomorphic, it suffices to check that $j_* L$ annihilates the defining relation for $Q$, namely we need
$$j_*L\rho=0.$$
However, this is precisely the statement that $L$ annihilates $t$, since $t=\rho\circ j$.

\begin{nitem}  The triple $(\C^3,Q,N)$ is a one-sided relative embedding.
\end{nitem}
Indeed, the embedding $j$ of $E$ into $\C^3$ is holomorphic, and $C^\infty$ up to the boundary.  Also, $N=j(E)\cap Q\cap U$ (for some open set $U$).  So $j(E)$ is a one-sided holomorphic extension of $N$ into the interior of $Q$.  

\begin{nitem}  The one-sided relative embedding is essential.
\end{nitem}

Assume that $N$ had a two-sided holomorphic extension $Z$, say.  Since $du$ and $dw$ are linearly independent on restriction to $N$, they remain linearly independent on $Z$ locally in a neighborhood of $p$.  By the implicit function theorem, $Z$ may be given locally as a graph $z=q(u,w)$ with $q$ holomorphic in a polydisc $\Delta_r((1,0))$ for polyradius $r=(r,r)$ sufficiently small.  Consider the function $wg(u)-q(u,w)$ in $\Delta_r((1,0))$.  This function vanishes identically on the hypersurface $T\cap\Delta_r((1,0))$.  But $T$ was proven to be pseudoconvex.  Therefore by the Lewy \cite{hl2} extension theorem, $wg(u)-q(u,w)$ vanishes identically in $E\cap\Delta_r((1,0))$ (after possibly shrinking $r$ if necessary).  But, by the choice of $g$, for all integers $k$ we have
$$\lim_{u\rightarrow 1} {g(u)\over (1-u)^k}=0.$$
It follows that $q(u,w)$ and all of its derivatives must vanish at $(1,0)$, which implies that $q(u,w)$ is identically zero in $\Delta_r((0,1))$.  Thus $g$ vanishes identically in a neighborhood of $u=1$, which is contrary to the choice of $g$. 
\end{proof}

The proof of Theorem \ref{essential} is now complete.

{\bf Remark.}  The example constructed in this section fails to be two-sided embeddable {\it at one point only}.  Ideally, one should be able to construct an example of a CR submanifold $N$ of the hyperquadric which fails to be two-sided embeddable at {\it all} of its points, perhaps by using a method analogous to that of Lewy \cite{hl1}.  However, it is not known whether such a construction is possible.

\section{The relative extension theorem}\label{relativesection}

\begin{lemma}\label{godown}  Suppose that $N$ is an abstract CR manifold of hypersurface type, and dimension $2k-1$.  Suppose also that there exists a CR embedding $\Phi:N\rightarrow \C^n$.  Then there exists locally a CR embedding $\Psi:N\rightarrow \C^k$ and a holomorphic submersion $\pi:\C^n\rightarrow\C^k$ such that the following diagram commutes:
$$
\xymatrix{
N \ar@{^{(}->}[r]^{\Phi} \ar@{^{(}->}[d]^{\Psi}& \C^n\ar[ld] \\ 
\C^k        &          \\ 
}
$$
\end{lemma}

\begin{proof}  We work locally at a point of $\Phi(N)$, which we shall choose to be the origin of $\C^n$, by a suitable translation.  By the rank theorem, it is possible to choose coordinates 
$$(\zeta,z,w)\in \C^{n-k}\times\C\times\C^{k-1}=\C^n$$
such that $\Phi(N)$ may be written as a graph
$$\Phi(N)=\{(\zeta,z,w)\in\C^n=\C^{n-k}\times\C\times\C^{k-1} : \zeta=G(x,w), y=g(x,w)\}$$
(where $z=x+iy$) with $G$ and $g$ smooth functions defined on $\R\times\C^{k-1}$ which vanish at the origin, together with their first partials.  (These latter two statements are furnished by taking a suitable affine change of the coordinates).
 
Thus, we may define a hypersurface in $\C^k$, $N_0$ say, by
$$N_0=\{(z,w)\in\C\times\C^{k-1} : y=g(x,w)\}.$$
Now, $N$ may be written as a graph over $N_0$ in the obvious way:
$$p=(x+ig(x,w),w)\mapsto \Psi(p)=(G(x,w),x+ig(x,w),w).$$
Define the mapping $\pi$ by $\pi(\zeta,z,w)=(z,w)$.  Then $\pi|_{\Phi(N)}$ is the inverse of $\Psi$.  But $\pi$ itself is holomorphic, so that $\Psi$ and $\pi|_{\Phi(N)}$ are CR isomorphisms onto their respective images.
\end{proof}

Now suppose given a CR manifold $Q$ embedded as a hypersurface in $\C^n$ with nondegenerate Levi form of signature $(k,l)$, with $k\ge 2$, $l\ge 1$, in a neighborhood of a point $x_0\in Q$.  Suppose also that we have a $(2k-1)$-dimensional strictly pseudoconvex CR manifold $N$ of hypersurface type which is CR embedded in $Q$, with $x_0$ lying in $N$.  Then the following lemma holds':

\begin{lemma}\label{locallewy}  Near $x_0$, $N$ has a one-sided extension into $\C^n$.  The direction of the extension is the direction of the concavity of $N$.
\end{lemma}

\begin{proof} $N$ is given as a CR submanifold of $\C^n$.  As in Lemma \ref{godown}, we may choose a mapping $\pi$ and an embedding $\Psi$ such that the diagram commutes
$$
\xymatrix{
N \ar@{^{(}->}[r]^{\Phi}\ar@{^{(}->}[d]^{\Psi}& \C^n\ar[ld]_\pi \\ %}  
\C^k        &          \\ 
}
.$$
In particular, there exist holomorphic coordinate functions $(z^1,...,z^n)$ on $\C^n$ and $(w^1,...,w^k)$ on $\C^k$ with
$$z^i=w^i\circ\pi\text{\ \ \ for\ }i=1,...,k.$$
Also, for $i=k+1,...,n$, the functions $z^i$ can be pulled through the diagram to define functions on $\Psi(N)$.  These functions will be CR on $\Psi(N)$, because the $z^i$ are holomorphic.  But, $\Psi(N)$ is a strictly pseudoconvex CR submanifold of $\C^k$ near $\Psi(x_0)$.  Hence, by the Lewy extension theorem, each $z^i$, for $i=k+1,...,n$, extends to a holomorphic function on the side of the concavity of $\Psi(N)$, say $\tilde{z}^i$.  Now $z^i=\tilde{z}^i$ defines a holomorphic graph in $\C^n$ over the concave side of $\Psi(N)$, which gives the one-sided extension.
\end{proof}

It should be remarked that the given local extension is {\it unique} (or at least its germ at a point of $N$ is unique), for if two extensions were given, then the difference of their graphing functions must be zero on $\Phi(N)$, and thus identically zero everywhere by the Lewy extension theorem.

\begin{theorem}\label{relativelewy}
$N$ has a global one-sided extension.  Furthermore, any two such extensions agree on another possibly smaller extension.
\end{theorem}
\begin{proof}  $N$ may be covered by a locally finite system $\{U_\alpha\}$ of neighborhoods, such that each $U_\alpha$ admits a one-sided extension (by the lemma).  Because of the uniqueness of local one-sided extensions, these local one-sided extensions must agree on the overlaps $U_\alpha\cap U_\beta$, and so patch together to give a global extension.
\end{proof}

The same set of arguments, only using the two-sided Lewy extension theorem, prove

\begin{theorem}  If $N$ is Levi nondegenerate and indefinite (instead of pseudoconvex), then near $x_0$, $N$ has a two-sided extension into $\C^n$.
\end{theorem}

\section{Twistor space}\label{twistorsection}

The purpose of this section is to review some standard facts about twistor space (in the flat case), and to fix notation for later sections.  More detailed accounts can be found in \cite{PenroseRindler2} or \cite{HuggettTod}.

Let $\TT$ be a $4$-dimensional complex vector space (called twistor space) equipped with the standard representation of $SL(4,\C)$.  One of the basic constructions of twistor theory is to exploit the isomorphism of $SL(4,\C)$ with $\operatorname{Spin}(6,\C)$, the simply connected cover of the conformal group in four dimensions in order to express data on Minkowski space in terms of data on twistor space and vice-versa.  Concretely, the $SL(4,\C)$ structure on $\TT$ is realized by a nonzero element $\epsilon\in\wedge^4\TT^*$.  This, in turn, defines a complex bilinear form on $\wedge^2\TT$ via $q(X,Y) = \epsilon(X\wedge Y)$, thus realizing the isomorphism $SL(4,\C)\cong Spin(6,\C)$. In the projectivization, $\PP(\wedge^2\TT)$, the zero locus of $q$ is none other than the Klein quadric, and is identified with the Grassmannian of $2$-planes in $\TT$, $\operatorname{Gr}_2(\TT)$.  The metric associated to $q$ is degenerate on $q=0$, but only in the complex scaling direction (along which $q$ is preserved up to scale).  So $q$ defines a complex conformal structure on $\operatorname{Gr}_2(\TT)$.  The quadratic form $q$ is conformally flat in any affine slice of the Klein quadric, so $\operatorname{Gr}_2(\TT)$ is locally conformally flat.   In the setting of interest, $\operatorname{Gr}_2(\TT)$ is called conformal compactified complexified Minkowski space, and is denoted $\widehat{\MM}_\C$.

By definition of the quadratic form $q$, two points $X,Y\in \PP(\wedge^2\TT)$ are null-related if and only if $X\wedge Y=0$.  In particular, two points of $\widehat{\MM}_\C$ are null-related if and only if the corresponding two-planes in $\TT$ share a line in common.  This incidence relation allows $\PP(\TT)$ to be recovered from $\widehat{\MM}_\C$ as one of the two (topologically distinct) sets of completely null $2$-planes in $\widehat{\MM}_\C$: the $\alpha$-planes are those whose annihilator at each point of $\mathbb{M}$ is an antiselfdual two-form on $\mathbb{M}$, and the $\beta$-planes are those annihilated by a selfdual two-form.

The tautological bundle of $\operatorname{Gr}_2(\TT)$ is the subbundle of $\TT\times \operatorname{Gr}_2(\TT)$ whose fibre over a point $x\in\operatorname{Gr}_2(\TT)$ is the $2$-plane in $\TT$ defined by $x$.  This bundle is denoted by $\overline{S}^*$, and is called the primed cospin bundle.  Let $S$ be the quotient bundle of $\TT\times\operatorname{Gr}_2(\TT)$ by $\overline{S}^*$.  This is the spin bundle.  An essential fact is:
\begin{itemize}
\item The tangent bundle of $\widehat{\MM}_\C$ is naturally identified with $S\otimes\overline{S}$.
\end{itemize}
Indeed, more generally if $\operatorname{Taut}$ denotes the tautological bundle for the Grassmannian $\operatorname{Gr}_k(\mathbb{C}^n)$ and $\operatorname{Taut}^{\operatorname{op}}$ is the quotient $(\C^n\times\operatorname{Gr}_k(\mathbb{C}^n))/\operatorname{Taut}$, then there is a natural isomorphism
$$T\operatorname{Gr}_k(\mathbb{C}^n) \cong \operatorname{Hom}(\operatorname{Taut},\operatorname{Taut}^{\operatorname{op}}).$$

Breaking the conformal invariance can be achieved by selecting a point $I\in\widehat{\MM}_\C$.  Then the metric, depending on $X\in\widehat{\MM}_\C$ away from the null cone of $I$, given by $g_X=q(dX,dX)/q(I,X)^2$ is flat.  The stabilizer of $I$ is the Poincar\'{e} group plus dilations (the semidirect product $S(GL(2,\C)\times GL(2,\C))\ltimes \mathbb{C}^4$).

Real Minkowski space can also be described from the twistor point of view.  Here we introduce a $(2,2)$ signature hermitian form $h$ on $\TT$ that is compatible with the $SL(4,\C)$ structure in the sense that $h(\epsilon,\epsilon)=24$.  This gives a reduction of the structure group to $SU(2,2)$.  The idea is then to exploit the isomorphism of $SU(2,2)$ with the spin group $\operatorname{Spin}(2,4)$ associated to the conformal group of a real Lorentzian metric in four dimensions (via a $4:1$ cover).  To this end, let $\NN$ be the null cone of $h$ in $\TT$.  The space of complex two-planes lying in $\NN$, denoted $\widehat{\MM}$ is compactified real Minkowski space.  It is a $4$-dimensional real submanifold of $\widehat{\MM}_\C$, equipped with the restriction metric.

The process of associating the real Minkowski space to the real quadric $\NN$ is also reversible. In the real case, the real hypersurface $\PP\NN\subset\PP\TT$ is a CR manifold with Levi signature $(1,1)$.  It can be identified with the space of null geodesics in $\widehat{\MM}$.  The projective spheres associated to the null cones at each point in $\widehat{\MM}$ each carry a canonical complex structure.  The CR structure on $\PP\NN$ is the unique CR structure such that the image of each of these spheres under the null geodesic spray is a holomorphic curve.  The rest of the section is devoted to establishing this fact.

\subsection{Complex structure on the projective sphere}\label{ComplexStructure}
Let $\mathcal{V}$ be an oriented four-dimensional real vector space, and let $H$ be the vector field generating the dilations on $\mathcal{V}$.  Suppose that $\mathcal{V}$ is equipped with a Lorentzian structure whose null cone is the set $\mathcal{N}\subset\mathcal{V}$.  We shall here exhibit a complex structure on conformal sphere $S^2$ obtained by quotienting $\mathcal{N}$ by the dilations.  

The Lorentzian structure on $\mathcal{V}$ gives rise to a metric $g_{\mathcal{V}}$ on $T\mathcal{V}$.  The orientation and metric together determine a $4$-vector $\sigma\in\wedge^4T\mathcal{V}$.  Define the duality operator $\star:\wedge^kT\mathcal{V}\to\wedge^{4-k}T\mathcal{V}$ by
$$(\star {\mathbf A})\wedge {\mathbf B} = g_{\wedge^k\mathcal{V}}({\mathbf A},{\mathbf B})\,\sigma$$
for all ${\mathbf B}\in\wedge^kM$.  The $g_{\wedge^k\mathcal{V}}$ appearing on the right hand side is the induced metric on multivectors, which on simple wedge products is given by
$$g_{\wedge^kg_{\mathcal{V}}}(A_1\wedge\cdots\wedge A_k,B_1\wedge\cdots\wedge B_K) = \det \left[g_{\mathcal{V}}(A_i,B_j)\right]_{1\le i,j\le k}.$$
The duality operator satisfies $\star^2=-(-1)^{k(k-4)}\operatorname{Id}$ on $\wedge^kT\mathcal{V}$.

Let $H^\perp$ be the orthogonal complement of the vector field $H$ with respect to $g_{\mathcal{V}}$.  Restricting to $\mathcal{N}$, $H\in H^\perp$, so we may form the quotient vector space $\mathcal{E}=H^\perp/\operatorname{span}(H)$.  We shall exhibit a line subbundle $\mathcal{M}$ of $\mathcal{E}\otimes\mathbb{C}$ over $\mathcal{N}$ that descends to the quotient space and gives a complex structure there.

Observe that wedging with $H$ yields an isomorphism of $\mathcal{E}$ with $H\wedge H^\perp\subset\wedge^2T\mathcal{N}$.

\begin{lemma}
The duality operator preserves $H\wedge H^\perp$:
$$\star(H\wedge H^\perp) = H\wedge H^\perp.$$
\end{lemma} 
\begin{proof}
The tangent space $\wedge^2T\mathcal{V}$ carries two nondegenerate pairings, that are related by $\star$:
\begin{align*}
(\mathbf A,\mathbf B) &\mapsto \mathbf{A}\wedge\mathbf{B}\\
(\mathbf A,\mathbf B) &\mapsto g_{\wedge^2\mathcal{V}}(\mathbf{A},\mathbf{B}).
\end{align*}
The orthogonal complement of $H\wedge H^\perp$ is equal to $H^\perp\wedge H^\perp + H\wedge T\mathcal{V}$ with respect to {\em both} of these pairings, and therefore the duality operator preserves $H\wedge H^\perp$.
\end{proof}

Since the duality operator preserves the two-dimensional $H\wedge H^\perp$ and $\star^2=-\operatorname{Id}$, there is a line subbundle $\mathcal{M}\subset\mathcal{E}\otimes\mathbb{C}$ such that $H\wedge \mathcal{M}$ is an eigenspace for the duality operator with eigenvalue $-i$.  Since self-duality is preserved by the group of dilations of $\mathcal{N}$, $\mathcal{M}$ descends to a complex line bundle on the quotient space $S^2$.  This line bundle is naturally identified with a subbundle of $TS^2\otimes\mathbb{C}$ which, in turn, defines a complex structure.

It is convenient to have a section of $\mathcal{M}$ in coordinates.  Suppose that $\mathcal{V}$ carries linear coordinates $(v^0,v^1,v^2,v^3)$ in terms of which the Lorentzian quadratic form is
\begin{equation}\label{Lorentzian}
(v^0)^2-(v^1)^2-(v^2)^2-(v^3)^2,
\end{equation}
and the metric is therefore $g_{\mathcal{V}}=(dv^0)^2-(dv^1)^2-(dv^2)^2-(dv^3)^2$.  Then, modulo $H$ and up to an overall scaling, a section of $\mathcal{M}$ has the form
\begin{equation}\label{M}
M = \left(v^2\frac{\partial}{\partial v^1}-v^1\frac{\partial}{\partial v^2}\right)+i\left(v^3\frac{\partial}{\partial v^0}+v^0\frac{\partial}{\partial v^3}\right).
\end{equation}

\subsection{The tangent bundle and null geodesic spray}
Let $\mathbb M$ be a four-dimensional manifold and $T\mathbb M'$ its tangent bundle with the zero section deleted.  Let $\pi:T\mathbb M'\to \mathbb M$ be the projection map.  Local coordinates $x^i$ defined on $\mathbb M$ induce fiber coordinates $v^i$ on $T\mathbb M'$ defined by writing a vector field $X$ in terms of the partials $\partial/\partial x^i$:
$$X = v^i(X)\frac{\partial}{\partial x^i}.$$
The double tangent bundle $TT\mathbb M'$ is a bundle over $T\mathbb M'$ with a local basis of $2n$ vector fields $\partial/\partial x^i,\partial/\partial v^i$.  The vertical subspace of $TT\mathbb M'$, denoted by $\mathcal{V}T\mathbb M'$, is the kernel of $d\pi:TT\mathbb M'\to T\mathbb M'$.  This is locally spanned by the vector fields $\partial/\partial v^i$.

There is a canonical diffeomorphism-invariant endomorphism $\lambda : TT\mathbb M'\to TT\mathbb M'$ of the double tangent bundle that is the generator of translations in the fibers of $T\mathbb M$.  It is given in these local coordinates by
$$\lambda = dx^i\otimes\frac{\partial}{\partial v^i}.$$
Note that the image and kernel of $\lambda$ are both $\mathcal{V}T\mathbb M'$.  In particular $\lambda^2=0$.  The endomorphism $\lambda$ allows us to define differentiation up the vertical direction.  To wit, if $X\in TT\mathbb M'$ and $f$ is a function on $T\mathbb M'$, then define $D_X f = \mathscr{L}_{\lambda X}f$.  The other relevant diffeomorphism-invariant structure on $T\mathbb M'$ is the Euler vector field $H$, which is the generator of scaling in the fibers.  In local coordinates
$$H = v^i\frac{\partial}{\partial v^i}.$$

Suppose now that $\mathbb M$ is equipped with a Lorentzian metric $g$.  This defines a quadratic form $G$ on $T\mathbb M'$.  The zero locus of $G$, denoted by $\mathcal{N}$, is the fibration of null cones.  The one-form $\alpha$ defined by $\alpha(X)=D_XG$ is a symplectic potential on $T\mathbb M'$.  The geodesic spray is the Hamiltonian vector field $V$ defined by $V\lrcorner d\alpha = dG$.  Note that $\lambda V=H$.  The integral curves of $V$ project to affinely parametrized geodesics on $\mathbb M$.  Since $V$ is everywhere tangent to $\mathcal{N}$, geodesics that start on $\mathcal{N}$ must stay on $\mathcal{N}$.  These are the null geodesics, and the restriction of $V$ to $\mathcal{N}$ is the null geodesic spray.

Because the integral curves of $V$ are geodesics, in local coordinates we have
\begin{equation}\label{VGamma}
\qquad\qquad\qquad V = v^k\frac{\partial}{\partial x^k} - \Gamma_{ij}^k v^iv^j\frac{\partial}{\partial v^k}
\end{equation}
where $\Gamma_{ij}^k$ are the Christoffel symbols of the metric on $\mathbb M$.

\subsection{Twistor space}
Assume that $\mathbb M$ is also oriented.  Twistor space $\mathbb{\PP\NN}$ is the space of (unparametrized) null geodesics in $\mathbb M$, which is the quotient of $\mathcal{N}$ by the $2$-parameter group of diffeomorphisms generated by the null geodesic spray $V$ and Euler vector field $H$.  Let $\mathcal{M}$ be the complex line bundle constructed fiberwise in \S\ref{ComplexStructure}, and let $M$ be given by \eqref{M} in local coordinates.  If $\mathbb M$ is conformally flat, then the complex vector fields $M,[V,M]$ are in involution with $V$ and $H$, and therefore descend to a CR structure on the quotient.  Involutivity follows at once by using a local coordinate chart in which the metric has the form \eqref{Lorentzian}, and applying \eqref{VGamma} and \eqref{M}.

\subsection{The connection and horizontal vectors}
The Levi-Civita connection associated to the metric $g$ on $\mathbb M$ allows us to lift any $C^1$ curve in $\mathbb M$ through a point to a $C^1$ curve $\gamma(t)$ through any point $X\in TT\mathbb{M}'$ in the fiber over $x$ for small time, by parallel translation of $X$ along $\gamma$.  This is the horizontal lift of the curve to the tangent bundle.  The tangent vector to the lifted curve is a vector in $TT\mathbb M'$, and is called a horizontal vector.  A vector is horizontal if and only if it is in the kernel of the operator $P:TT\mathbb M'\to \mathcal{V}T\mathbb M'$ given by
$$P = \left(dv^k + dx^i\Gamma_{ij}^kv^j\right)\otimes\frac{\partial}{\partial v^k}$$
where $\Gamma_{ij}^k$ are the Christoffel symbols of the connection.  Invariantly, $P=\frac{1}{2}(\operatorname{Id}_{TTM'} + \mathscr{L}_V\lambda)$.  The space of horizontal vectors $\mathcal H=\ker P$ is a complementary subspace to $\mathcal VT\mathbb{M}'$:
$$TT\mathbb M' = \mathcal H\oplus\mathcal VT\mathbb M'.$$
The operator $P$ has image $\mathcal{V}T\mathbb{M}'$ and satisfies $P^2=P$, and so is the projection operator $P:TT\mathbb{M}'\to \mathcal{V}T\mathbb{M}'$ associated to this splitting.  Note in particular that the dynamical vector field $V$ is horizontal.

\subsection{Umbral bundle}
We describe here a vector bundle, known as the {\em umbral bundle} \cite{HSCausal}, on $\mathcal{N}$ that is needed to develop the theory of shear-free congruences.  The pullback bundle $\pi^{-1}T\mathbb M\cong T\mathcal{N}/\mathcal{V}\mathcal{N}$ is isomorphic via $\operatorname{Id}-P$ to the bundle $\mathcal H$ of horizontal vectors.  The metric $g$ on $\mathbb M$ lifts to a metric in the pullback bundle, and so defines a metric on horizontal vectors denoted by $g_{\mathcal H}$.  Along $\mathcal{N}$, $V$ is a null vector with respect to $g_{\mathcal{H}}$, and so $V$ lies in its own orthogonal complement $V^\perp \subset \mathcal H$.  Define the umbral bundle to be the vector bundle
$$E = \frac{V^\perp}{\operatorname{span}(V)}.$$
The bundle $\mathcal H\cong\pi^{-1}T\mathbb M$ inherits an orientation from $\mathbb M$.  The metric and orientation define a duality operator on $\wedge^2\mathcal H$ which, as in \S \ref{ComplexStructure}, preserves the subspace $V\wedge E$.  Select a horizontal complex vector field $M'$ in $V^\perp\otimes\mathbb{C}$ such that $\star(V\wedge M') = -i(V\wedge M')$.  This vector field is defined uniquely up to an overall factor, and modulo multiples of $V$.

\begin{lemma}
$M'\equiv 0 \pmod{V,\,M,\,[V,M]}$
\end{lemma}
\begin{proof}
The identity $M'=-[V,\lambda M']$ holds for $M'$ horizontal.  However, $\lambda M'$ is a vertical vector such that $H\wedge \lambda M'$ is self-dual, and therefore $\lambda M'$ is a section of $\mathcal{M}$.
\end{proof}

\subsection{Null geodesic congruences}\label{congruences}
A null geodesic congruence is a nonvanishing null geodesic vector field $k$ throughout a region of space time.  Equivalently, this is a smooth section $s_k:M\to \mathscr{N}$ of the null cone bundle over $\mathbb{M}$ whose image is foliated by the null geodesic spray. The pullback of the umbral bundle associated along $s_k$ is the bundle of $2$-planes $k^\perp/\operatorname{span}(k)$ consisting of vectors orthogonal to $k$, modulo multiples of $k$ itself (\cite{HSCausal}).  The physical significance of this bundle is (Sachs \cite{Sachs}) that if an object is held in the family of light rays generated by $k$, then the shadow (or umbra) it casts on a screen is naturally described on the umbral bundle.  Note that the metric descends to a negative definite form in $k^\perp/\operatorname{span}(k)$.

The umbral bundle associated to $s_k$ carries a natural complex structure, defined in a similar way to \S \eqref{ComplexStructure}.  Wedging with $k$ gives a natural linear isomorphism $k^\perp/k\cong k\wedge k^\perp$.  The duality operator preserves $k\wedge k^\perp$, since the orthogonal complement of $k\wedge k^\perp$ is $\wedge^2 k^\perp + k\wedge T\mathbb M$ and this also annihilates $k\wedge k^\perp$ under the wedge product.  Since $\star^2=-1$, the eigenspaces of $\star$ define a complex structure on $k\wedge k^\perp$ and therefore also on $k^\perp/k$.  Let $m$ be a null section of $k^\perp\otimes\mathbb{C}$ such that the coset of $m$ modulo $k$ is an element of the $+i$ eigenspace of the star operator, scaled so that $g(m,\overline{m})=-1$.  Now, there exists a unique null vector field $n$ such that $g(m,n)=0$, $g(k,n)=1$.  The quadruple of null vectors $(k,n,m,\overline{m})$ is a {\em null tetrad}.

The distortion tensor $\nabla_a k_b$ descends to a bilinear form on $k^\perp/k$.  The trace-free symmetric part is called the {\em shear} of the congruence.  The skew part is called the {\em rotation}.

\begin{theorem}\label{SF=>CR}
A shear-free null geodesic congruence defines a CR submanifold of $\PP\N$.  The Levi form of this CR manifold is the rotation of the congruence.
\end{theorem}
\begin{proof}
Complete the vector $k$ to a null tetrad basis $k,n,m,\overline{m}$ such that the only non-vanishing inner products are $k\cdot n = n\cdot k = - m\cdot\overline{m} = - \overline{m}\cdot m = 1$ and with $k \wedge m$, $n \wedge \overline{m}$ and $k\wedge n - m \wedge \overline{m}$ all anti-self-dual and  $k \wedge\overline{m}$, $n \wedge m$ and $k\wedge n + m \wedge \overline{m}$ all self-dual.  In terms of the basis, the shear and rotation are the scalars
\begin{align*}
\sigma &= (\mathscr{L}_kg)(m,m)=-2g([k,m],m)\\
\rho &= idk^\flat(m,\overline{m})=ig([m,\overline{m}],k).
\end{align*}

The vector field $m$ descends up to scale to the three-dimensional space of geodesics that form the congruence if and only if $[k,m]\wedge k\wedge m=0$.  Write $[k,m] = g([k,m],k)n + g([k,m],n)k - g([k,m],m)\overline{m} - g([k,m],\overline{m})m$.  Since $k$ is a null geodesic and $m$ is orthogonal to $k$, it follows that $g([k,m],k)=0$. Hence $[k,m]\wedge k\wedge m = -g([k,m],m)\overline{m}\wedge k\wedge m$.  Thus $k$ is shear free if and only if $m$ descends to a complex direction field on the quotient space.  This generates a CR structure.  The Levi form of the CR structure is given by $i[m,\overline{m}]\pmod{k,m,\overline{m}}$.  Writing $[m,\overline{m}]$ in the tetrad, the Levi form is $ig([m,\overline{m}],k)=\rho$, the rotation of the congruence.

It remains to show that the CR structure of a shear free congruence is compatible with the CR structure on twistor space.  The null vector field $k$ defines a section $s:M\to\mathcal{N}$ of the null cone bundle.  The null geodesic spray is tangent to this section, since $k$ is a geodesic, and $s_*k=V$.  The vector $s_*m$ is in $V^\perp$, and satisfies $\star\!(V\wedge s_*m) = -i(V\wedge s_*m)$.  Because this property also characterizes $M'$, $s_*m \equiv 0 \pmod{\mathcal{V}T\mathbb{M}, V, H, M'}$.  It is therefore sufficient to show that $Ps_*m\equiv 0\pmod{H,M}$.  It is shown in \cite{HSCausal} that $Ps_*m = \lambda s_*\nabla_mk$.  This is a linear combination of $H=\lambda s_*k$ and $M=\lambda s_*m$ if and only if $\nabla_mk$ is a linear combination of $k$ and $m$, if and only if $k$ is shear free.
\end{proof}

For the converse, let $N$ be a CR submanifold of $\PP\NN$.  Define a {\it cone over $N$} to be an open family $\hat{U}$ of lines in $\PP\NN$ such that every point of $N$ lies in some line and such that every line in $\hat{U}$ contains some point of $N$.  Note that in light of what has been said before, $\hat{U}$ defines an open subset $U$ of Minkowski space.

\begin{theorem}\label{CR=>SF}  Let $N$ be a CR submanifold of $\PP\NN$ and let $\hat{U}$ be a cone over $N$.  Suppose that every member of $\hat{U}$ which intersects $N$ does so in a unique point.  Then there is a shear-free congruence $k^a$ on $U$ of which $N$ is the associated CR manifold.
\end{theorem}
\begin{proof}  The intersection of each line with $N$ defines a smooth null vector field $k$ in an open subset $U$ of Minkowski space.  The CR structure defines (up to scale) a complex vector field $m$ in $U$ that is Lie derived along $k$ and is orthogonal to $k$.  Hence $k$ is shear-free (hence also geodesic).  By construction, $N$ is the CR manifold associated to $k^a$.
\end{proof}

Given a point $p$ in $N$, there is an open neighborhood $W$ of $p$ and a cone $\hat{U}$ over $N\cap W$ such that the hypothesis of the theorem is satisfied.  Therefore, the theorem holds locally for any CR manifold $N$.  The requirement that every line meeting $N$ does so in a unique point is tantamount to requiring that the congruence be {\it single-valued}.  There are important examples of congruences which are not single-valued, such as the Kerr congruence which is a congruence with branch cuts.  The techniques employed in this paper do apply to such congruences locally, on any single-valued branch.

%%%%%%%%%%%%%%
\section{Non-analytic congruences}\label{nonanalytic}
Penrose \cite{PenroseAlg} suggested that a shear-free rotating congruence might necessarily be real-analytic.  In this section, we prove that some such congruences are non-analytic.  We need the following fact (attributed to Kerr in \cite{PenroseAlg}) about the relationship between the CR submanifold and the structure of the congruence:

\begin{lemma}  The following statements are equivalent:\\
\noindent 1.  $N$ is the intersection of $\PP\N$ with a complex-analytic submanifold of $\C\PP^3$.\\
\noindent 2.  The associated congruence is a real-analytic shear-free congruence of null geodesics.
\end{lemma}
\begin{proof} A real-analytic congruence is by definition a null geodesic whose tangent vector field $k$ is a real-analytic function of the coordinates of Minkowski space in some open set $U$ in the real Minkowski space.  We may thus extend $k$ to a complex-analytic null vector field on a neighborhood of $U$ in the complexified Minkowski space.  When $k$ is descended to the leaves of the twistor foliation, we obtain a complex analytic submanifold of $\PP\TT$, as required.  The converse follows by reversing the line of argument.
\end{proof}

Applying the lemma, Theorem \ref{essential} assumes the following form:

\begin{theorem}\label{antirob}  There exist nonanalytic congruences of null geodesics whose shear vanishes, and whose rotation is non-zero.
\end{theorem}

\section{Null electromagnetic fields}\label{nullfields}
A $2$-form $F$ on an open set in $\mathbb{M}$ is a source-free (or homogeneous) electromagnetic field if the following field equations hold:
$$ dF =0, \qquad d\star\!\!F =0$$
where $\star$ is the duality operation associated to the metric.  If we let $G=F+i\star\!\!F$, then the equations assume the simple form
$$dG=0,$$
and since $F$ is real, $G$ determines $F$ completely.  In this language, solutions of the homogeneous electromagnetic field equations correspond to complex two-forms $G$ that obey
$$dG=0$$
and which are self-dual ($\star\!G=-iG$).  We concentrate on the case where the field tensor is null, i.e.,
$$g_{\wedge^2T\mathbb{M}}(F,F)=g_{\wedge^2T\mathbb{M}}(F,\star\!F)=0,$$
or in terms of $G$,
$$g_{\wedge^2T\mathbb{M}}(G,G)=0.$$
A field $G$ satisfying this equation is also called {\em null}.  Thus $G$ is null if and only if $F$ is null.

\begin{lemma}\label{TypeNShearfree}
Let $G$ be a non-zero complex self-dual null $2$-form.  Then $G=k^\flat\wedge m^\flat$ where $k$ and $m$ are mutually orthogonal null vectors with $k$ real.  If, in addition, $G$ is closed, then $k$ is tangent to a shear-free null geodesic congruence.
\end{lemma}

A field $G$ of the form $G=k^\flat\wedge m^\flat$, as in the lemma, is called {\em adapted to $k$}.  Thus $G$ is adapted to $k$ if and only if the associated Maxwell field $F$ is itself null and adapted to $k$.

\begin{proof}
Since $G$ is self-dual, $g_{\wedge^2T\mathbb{M}}(G,G)=0$ implies that $G\wedge G=0$, so $G$ splits as an outer product $G=u^\flat\wedge v^\flat$ for some (complex) linearly independent vectors $u$ and $v$.  Now, $u^\flat\wedge G=0$, so again by self-duality $u\lrcorner G=0$, and so $u$ is null and orthogonal to $v$.  Likewise, $v$ must also be null.  Next, we claim that at any point where $G$ is not zero there is a real vector $k$, necessarily null, in the linear span of $u$ and $v$.  Since $G$ is self-dual, $\overline{G}$ is antiself-dual, so $G\wedge\overline{G}=0$, from which the claim follows.

Now suppose that $dG=0$.  It follows that $\mathscr{L}_kG=0$.  Using the decomposition $G=k^\flat\wedge m^\flat$, this implies $(\mathscr{L}_kk^\flat)\wedge m^\flat\wedge k^\flat=0$, or $\mathscr{L}_kk^\flat \equiv 0\pmod{k^\flat, m^\flat}$.  But, since $\mathscr{L}_kk^\flat$ is real, this implies that $\mathscr{L}_kk^\flat=\alpha k^\flat$ for some $\alpha$.  Now because $k$ is null, $\mathscr{L}_kk^\flat = \nabla_kk^\flat$.  So $\nabla_kk^\flat=\alpha k^\flat$.  The constant $\alpha$ can be absorbed by rescaling $k$ (and so also rescaling $m$), so that $\nabla_kk^\flat=0$, and thus $k$ is a null geodesic.  Finally,
$$0=m^\flat\wedge\mathscr{L}_mG = -k^\flat\wedge m^\flat\wedge \nabla_mm^\flat$$
Applying duality gives $0=\langle k,\nabla_mm\rangle=\frac{\sigma}{2}$.
\end{proof}

\section{Local existence}\label{localexistencesection}
The classical theorem on local existence of solutions of the null field equations for flat space-time and {\em analytic} congruences is due to Robinson \cite{Robinson} and Sommers \cite{Sommers}:
\begin{theorem}  Let $k$ be vector field tangent to an analytic shear-free congruence of null geodesics.  Then there exists locally a solution of the electromagnetic field equations $F$ that is adapted to $k$.  The freedom in the solution is described by a holomorphic function of two complex variables.
\end{theorem}

We may drop the assumption of analyticity of the congruence, provided that we assume that the congruence is everywhere rotating.  Let $k$ be a shear-free congruence.  Associated to $k$ is the CR manifold $N\subseteq\PP\NN$.

\begin{lemma}\label{max=db}  Let $k$ be a shear-free congruence.  Let $G$ be a self-dual nonvanishing 2-form.  The following conditions are equivalent:

1.  $G$ is the pullback of a $\db$-closed $(2,0)$ form on $N$ to the folia of congruence generated by $k$.

2.  $G$ is a null solution of the electromagnetic field equation and is adapted to $k$.
\end{lemma}

\begin{proof}  Locally, $N$ is the quotient of an open set $U$ in $\mathbb{M}$ by the leaves of the foliation defined by $k$ via a submersion $\phi:U\to N$.

Suppose that $H$ is a given $\db$-closed $(2,0)$ form on $N$, and let $G=\phi^*H$.  If $H$ is closed, then so is $G$. Recall the definition of $\db:Q^{2,0}\rightarrow Q^{2,1}$ where
$$Q^{2,0}=\Omega^{1,0}\wedge\Omega^{1,0}$$
and
$$Q^{2,1}={\Omega^{1,0}\wedge\Omega^{1,0}\wedge\Omega^1\over \Omega^{1,0}\wedge\Omega^{1,0}\wedge\Omega^{1,0}}.$$
The operator $\db$ is the induced mapping of $d$ on the indicated quotient spaces.  However, in the above expression for $Q^{2,1}$, the space $\Omega^{1,0}$ is two-dimensional, and so its triple exterior product vanishes.  Thus
$$Q^{2,1}=\Omega^{1,0}\wedge\Omega^{1,0}\wedge\Omega^1$$
and
$$d^{2,0}=\db^{2,0}.$$
Hence $dH=0$.  So it follows that $G=\phi^*H$ is closed.  Now $k^\flat$ and $m^\flat$ descend to a basis of $\Omega^{1,0}$, so $G$ must have the form $\mu k^\flat\wedge m^\flat$.  So condition 1 implies condition 2.

Conversely, if $G=\mu k^\flat\wedge m^\flat$ is a closed form, then $\mathscr{L}_kG=i_kG=0$, and therefore $G$ descends to a closed $(2,0)$-form $H$ on $N$.
\end{proof}

Assume now that $k$ is rotating.  Then $N$ is strictly pseudoconvex, and so there is a one-sided holomorphic extension $Z$.  Now consider the sheaf of $\bar{\dd}$-closed $(2,0)$-forms on $Z$ which are $C^\infty$ up to the boundary $N$.  Any section of this sheaf in a neighborhood of a point on $N$ will induce a local solution of the electromagnetic field equations on $N$.  Since $Z$ can be locally embedded into $\C^2$, any local holomorphic section of the canonical line bundle of $\C^2$ will induce a section of the aforementioned sheaf by restriction, and thus a local solution of the electromagnetic field equations which is null with respect to $k^a$.  

If now $G_1$ and $G_2$ are two nonzero null self-dual electromagnetic fields adapted to $k$, then the associated $(2,0)$-forms $H_1$ and $H_2$ on $N$ satisfy $H_2=fH_1$ for some CR function $f$ on $N$.  By Lewy's extension theorem, $f$ can be extended on one side to a holomorphic function on the complex surface $Z$ whose boundary in $N$.

To summarize, we have:

\begin{theorem}\label{localexistence}  If $k$ is a shear-free rotating congruence, then there exists locally a solution of the electromagnetic field equations which is null with respect to $k$.  The freedom in the solution is described by the boundary value of a holomorphic function of two variables. A global solution exists if and only if there exists a global $\db$-closed $(2,0)$-form on the CR manifold $N$ associated with the congruence.
\end{theorem}

%The question of existence is treated in a somewhat different context in \cite{Tafel}.

\bibliographystyle{plain}
\bibliography{null_electromagnetic_fields}

\end{document}